\documentstyle[epsfig,color,axodraw,12pt]{article}

\def\plb#1{Phys.~Lett.~{\bf B#1}}
\def\npb#1{Nucl.~Phys.~{\bf B#1}}
\def\prl#1{Phys.~Rev.~Lett.~{\bf #1}}
\def\prd#1{Phys.~Rev.~{\bf D#1}}

\def\mt{{\ifmmode M^{eff}_T\else $M^{eff}_T$\fi}}

\def\e{\epsilon}

\def\l{\left}
\def\r{\right}

\def\ln#1{\mbox{ln}\l(#1\r)}

\def\e3{$\epsilon_3$}

\def\ch2{$\chi^2$}

\def\co#1{{\ifmmode{\cal O}_{#1}\else${\cal O}_{#1}$\fi}}

\newdimen\unit
\def\point#1 #2 #3{\vbox to0pt{\kern-#2\unit
  \hbox{\kern#1\unit#3}\vss}
 \nointerlineskip}

\newcommand{\be}{\begin{equation}}
\newcommand{\ee}{\end{equation}}
\newcommand{\bea}{\begin{eqnarray}}
\newcommand{\eea}{\end{eqnarray}}

\begin{document}
\thispagestyle{empty} \noindent
\begin{flushright}
        April 2003 \\
        OHSTPY-HEP-T-03-005
\end{flushright}

\vspace{1cm}
\begin{center}
Phenomenology of the Minimal SO(10) SUSY Model\footnote{Invited
talk at PASCOS'03, Mumbai, India, January 3 - 8, 2003.  This talk
is based on work in collaboration with T. Bla\v{z}ek, R. Derm\' \i
\v sek, L. Roszkowski, R. Ruiz de Austri and K. Tobe}
\end{center}
  \vspace{1cm}
    \begin{center}
Stuart Raby\\
      \vspace{0.3cm}
\begin{it}
Department of Physics, The Ohio State University, \\ 174 W. 18th
Ave., Columbus, OH 43210; \\ On leave of absence, School of
Natural Sciences, \\ Institute for Advanced Study, Princeton, NJ
08450
\end{it}
  \end{center}
  \vspace{3cm}
\centerline{\bf Abstract}
\begin{quotation}
\noindent  In this talk I define what I call the minimal SO(10)
SUSY model. I then discuss the phenomenological consequences of
this theory, vis a vis gauge and Yukawa coupling unification,
Higgs and super-particle masses, the anomalous magnetic moment of
the muon, the decay $B_s \rightarrow \mu^+ \ \mu^-$ and dark
matter.
\end{quotation}
\vfill\eject

\section{Minimal SO(10) SUSY Model}
 Let me first define the minimal SO(10) SUSY model
[MSO$_{10}$SM]~\cite{bdr} and then I will discuss the
phenomenological consequences of this theory.   In the
MSO$_{10}$SM the quarks and leptons of one family are contained in
a {\bf 16} dimensional spinor representation and the two Higgs
doublets of the MSSM come from a single {\bf 10} dimensional
representation. We have

$$ \bf 16 \;\; \supset  \left[  Q = \left(\begin{array}{c} u \\
 d \end{array}\right), \;\;\;
L= \left(\begin{array}{c} \nu \\
 e \end{array}\right), \;\;\;
\bar{u}, \;\;  \bar{d}, \;\;\; \bar{e}, \;\;\;  \bar \nu \right]$$

$$ \bf 10_H \hspace{.2in} \supset \hspace{.2in} [ H_u,\;\; H_d,\;
\;\;  T,\;\; \bar T]  $$

For the third generation, there is a unique Yukawa coupling to the
Higgs doublets with $$ W \supset \lambda \ {\bf 16_3 \ 10_H \
16_3} .$$ As a consequence,  the top, bottom, tau and $\nu_\tau$
Yukawa couplings satisfy Yukawa unification with  $\lambda_t \ = \
\lambda_b \ = \ \lambda_\tau \ = \ \lambda_{\bar \nu_\tau}  \
\equiv \ \lambda$.   Note with a large Majorana mass for $\bar
\nu_\tau$ we have a see-saw mechanism resulting in a light
left-handed neutrino, i.e. $ M_{\bar \nu} \ \bar \nu_\tau \ \bar
\nu_\tau$ $\Longrightarrow$  $\;\; m_{\nu_\tau} \sim {\Large
m_t^2/M_{\bar \nu}}$.   Although I will not discuss Yukawa terms
for the first and second generation of quarks and leptons, it is
well known that it is not phenomenologically acceptable for them
to receive all their mass via renormalizeable interactions with a
single ${\bf 10_H}$. Nevertheless with effective higher
dimensional interactions it is not difficult to obtain realistic
fermion masses and mixing angles for all quarks and
leptons~\cite{brt}. Moreover if these mass matrices are
hierarchical, we do not significantly affect the results derived
from assuming exact Yukawa unification for the third generation.

Finally, the soft SUSY breaking parameters are given by
 $ - {\cal L}_{soft} =  m_{16}^2  \ \Sigma_{i =1}^3 \
 16_i^* 16_i +  m_{10}^2 \  10_H^* 10_H - A_0 \
\lambda \  16_3  10_H  16_3 $ $ + M_{1/2} \ \Sigma_{i =1}^3 \ (
\chi_i \chi_i) + \mu B \  H_u H_d $.  All but one of these terms,
are the most general consistent with SO(10).  A universal scalar
mass $m_{16}$ for all three families is an additional assumption.
Hence, the soft SUSY breaking parameters are given by  $$m_{16}, \
m_{10},\ A_0, \ M_{1/2}, \ \tan\beta. $$   Before continuing we
note that one additional soft SUSY breaking parameter is needed,
which we discuss next.

\subsection{Radiative EWSB with large $tan\beta$ needs $m_{H_u}^2 <
m_{H_d}^2$}

It has been shown that there are two consequences of splitting the
two Higgs doublet masses.  It reduces the amount of fine tuning
for radiative electroweak symmetry breaking [EWSB]
~\cite{rattazzi}. In addition, it permits EWSB in an entirely new
region of SUSY parameter space with $m_{16} \gg M_{1/2}$
~\cite{ewsb}.

We have considered the possibility of both $D_X$ term and ``Just
So" splitting~\cite{bdr}.   In the former, we assume a soft SUSY
breaking D term where $D_X$ is the auxiliary field of a $U(1)_X$
gauge interaction defined by $SO(10) \rightarrow SU(5) \times
U(1)_X$. We then obtain \begin{eqnarray}  m_{(H_u,\; H_d)}^2 = &
m_{10}^2
\mp 2 D_X &  \nonumber \\
 m_{(Q,\; \bar u,\; \bar e)}^2 = &  m_{16}^2 + D_X & \nonumber \\
 m_{(\bar d,\; L)}^2 = & m_{16}^2 - 3 D_X . &  \nonumber
 \end{eqnarray}
These boundary conditions at the GUT scale generically give the
low energy result $ m_{\tilde b}^2 \; \leq \;  m_{\tilde t}^2$
which is {\em bad} for Yukawa unification.

With ``Just So" splitting we have \begin{eqnarray}  m_{(H_u,\;
H_d)}^2 = & m_{10}^2 \; ( 1 \mp \Delta m_H^2) & \nonumber \\
 m_{(Q,\; \bar u,\; \bar e)}^2 =  & m_{16}^2  & \nonumber \\
 m_{(\bar d,\; L)}^2 = & m_{16}^2 . & \nonumber \end{eqnarray}
These boundary conditions give $ m_{\tilde t}^2 \; << \; m_{\tilde
b}^2$ which is  {\em good} for Yukawa unification.   This latter
case may be motivated by the fact that the Higgs multiplets must
be special.  They necessarily have a $\mu$ term and they also
require doublet-triplet splitting.   Moreover they have large
threshold corrections at the GUT scale due to the tau neutrino.

The $\bar \nu_\tau$ contribution to Higgs splitting results from
the Yukawa term ($\lambda_{\nu_\tau} \; \bar \nu_\tau \; L \;
H_u$) with $\lambda_{\nu_\tau} = \lambda_t = \lambda_b =
\lambda_\tau \equiv \; {\bf \lambda}$.  Since $\bar \nu_\tau$
couples only to $H_u$, this contribution distinguishes $H_u$ and
$H_d$.   At one loop we find $ \Delta m_{H_u}^2 \approx
\frac{\lambda^2}{16 \pi^2} \; (2 m_{16}^2 + m_{10}^2 + A_0^2) \log
(\frac{M_{\bar \nu_\tau}^2}{M_G^2}) + \; \cdots$.   Taking typical
GUT values for the parameters $\lambda = 0.7, \;$ $M_{\bar
\nu_\tau} = 10^{14} \; {\rm GeV}$ (which gives $(\Delta
m_\nu^2)_{atm} \sim 10^{-2} \ {\rm eV}^2$)$,\; M_G = 3 \times
10^{16} \; {\rm GeV}$  and $A_0^2 \approx 2 \, m_{10}^2 \approx 4
\, m_{16}^2$ we obtain $\Delta m_H^2$ $\equiv \frac{1}{2} \,
\Delta m_{H_u}^2/m_{10}^2 \sim .07$. This is ``Just So" splitting
of about the right size.

\subsection{Gauge coupling unification}

Presently, gauge coupling unification provides the only evidence
for low energy SUSY~\cite{susygut1,susygut2,susygutexp}.
\begin{center}
\SetPFont{Helvetica}{15}
%\SetScale{0.8}
\begin{picture}(300,45)(0,0)
\SetColor{Red} \Line(180,28)(50,15) \Line(180,25)(50,35)
\Line(180,25)(50,45) \Text(35,10)[1c]{$\alpha_3^{-1}$}
\Text(35,30)[1c]{$\alpha_2^{-1}$}
\Text(35,45)[1c]{$\alpha_1^{-1}$}

\SetColor{Black} \Line(180,26)(220,22)
\Text(227,30)[1c]{$\alpha_G^{-1}$} \Text(50,0)[1c]{$M_Z$}
\Text(180,0)[1c]{$M_G$}

\SetColor{Black}
\end{picture}
\end{center}
Note, when threshold corrections are included, the three gauge
couplings $\alpha_i, \; i = 1,2,3$ do not precisely meet at the
GUT scale.  Moreover, for consistency, one loop threshold
corrections need to be included when using two loop RG running
from $M_G \rightarrow M_Z$.  At one loop there are significant GUT
threshold corrections from Higgs and GUT breaking sectors.  We now
define the GUT scale as the point where $\alpha_1(M_G) =
\alpha_2(M_G) \equiv \tilde \alpha_G$.  A good fit to low energy
data then requires $\epsilon_3 \equiv \frac{(\alpha_3(M_G) -
\tilde \alpha_G)}{\tilde \alpha_G} \sim - 4\%$.

\section{SO(10) Yukawa unification}
Let us now consider the constraint on the soft SUSY breaking
parameters resulting from Yukawa unification~\cite{so10,bdr,tobe}.
Note, the GUT threshold corrections to Yukawa unification from
gauge and Higgs loops is typically insignificant.  Weak scale
threshold corrections, on the other hand, are proportional to
$\tan\beta$ and cannot be ignored~\cite{threshcorr,bpr}.  The
dominant contributions are given by $\delta m_b/m_b =  \Delta
m_b^{\tilde g} + \Delta m_b^{\tilde \chi} + \Delta m_b^{Log} +
\cdots $ where the first comes from a gluino-sbottom loop, the
second from the chargino-stop loop and the third from finite wave
function renormalization graphs.  Note, in general we have
 $\Delta m_b^{\tilde g} \sim - \Delta m_b^{\tilde
\chi} > 0$ for {\bf $\mu > 0$} [ our conventions ].  The first two
contributions are $\tan\beta$ enhanced and can be $\sim  50$ \%,
while the typical size of the log contribution is $ \sim + 6$ \%.
The contribution to the top quark mass is not $\tan\beta$ enhanced
and although the contribution to the tau mass is; nevertheless it
is small due to the smaller values of the relevant gauge and
Yukawa couplings at the electroweak scale.  Finally, good fits to
top, bottom and tau masses require $\delta m_b/m_b \leq -2$ %.

\subsection{Data favors $\mu > 0$}

We now argue that two pieces of low energy data favor positive
values of $\mu$. The first is the rate for the process $b
\rightarrow s \gamma$ and the second is the anomalous magnetic
moment of the muon.  In the first case, the chargino term
typically gives the dominant SUSY contribution and for $\mu > 0$
it has opposite sign to the standard model and charged Higgs
contributions, thus reducing the branching ratio.  This is
desirable since the SM contribution by itself is a little too
large.   As a result, trying to fit the data with $\mu < 0$ is
problematic. In the second case, the contribution to the muon
anomalous magnetic moment due to new physics (beyond the standard
model) is measured to be $a_\mu^{NEW} \times 10^{10}$ = 33.9
(11.2) [$e^+ \ e^-$ - based] or 16.7 (10.7) [$\tau$ -
based]~\cite{gminus2}. There are two results depending on whether
one uses $e^+ \ e^-$ or $\tau$ data to determine the hadronic
contribution to the amplitude. Note, in either case the sign of
$a_\mu^{NEW}$ is positive. Moreover in SUSY this sign is directly
correlated with the sign of $\mu$~\cite{nath}.  Again favoring
positive $\mu$. Hence we consider only positive $\mu$ in our
analysis.

\subsection{$\chi^2$ Analysis}
We have performed a $\chi^2$ analysis of the MSO$_{10}$SM with 11
input parameters defined at the GUT scale and 9 low energy
observables in our $\chi^2$ function~\cite{bdr}.

The 11 input parameters at $M_G$ are [$\lambda, \; \alpha_G, \;
M_G, \; \epsilon_3; \;$ $m_{10}, \; A_0, \; \tan\beta(M_Z), \;\;$
$D_X$ [ D term splitting]  ( or  $\Delta m_H^2$ [Just So Higgs
splitting]),$\;\; m_{16}, \; \mu, \; M_{1/2}$], where the last
three parameters are fixed while we vary 8 parameters using the
CERN package Minuit to minimize $\chi^2$.  The 9 observables
(experimental/theoretical uncertainty) [$X_i^{exp} \; (\sigma_i)$]
defining $ \chi^2 = \Sigma_{i = 1}^9 \left[ \frac{(X_i^{exp} -
X_i^{theory})^2}{\sigma_i^2} \right]$ are given by [$G_\mu, \;
\alpha,\; \alpha_s(M_Z) = 0.118 \;(0.002), \;$ $\rho^{NEW}, \;
M_Z, \; M_W, \;$    $M_t = 174.3 \;(5.1), \;$ $m_b(m_b) = 4.20
\;(0.20), \; M_\tau$].

\subsection{Bottom Line}

\begin{figure}[htbp]
\epsfxsize=15cm \centerline{\epsfbox{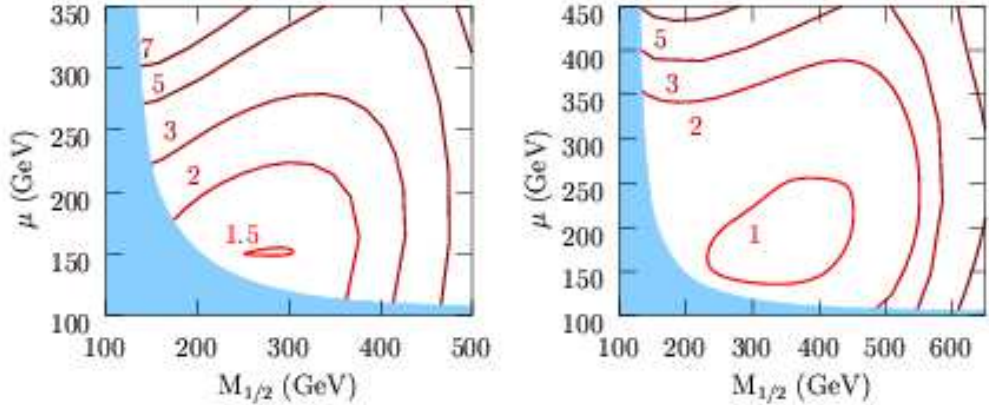}} \caption{$\chi^2$
contours for $m_{16} = 1500$ GeV (Left) and $m_{16} = 2000$ GeV
(Right).  The shaded region is excluded by the chargino mass limit
$m_{\chi^+} > 103$ GeV.} \label{fig:fig1}
\end{figure}

\begin{figure}[htbp]
\epsfxsize=15cm \centerline{\epsfbox{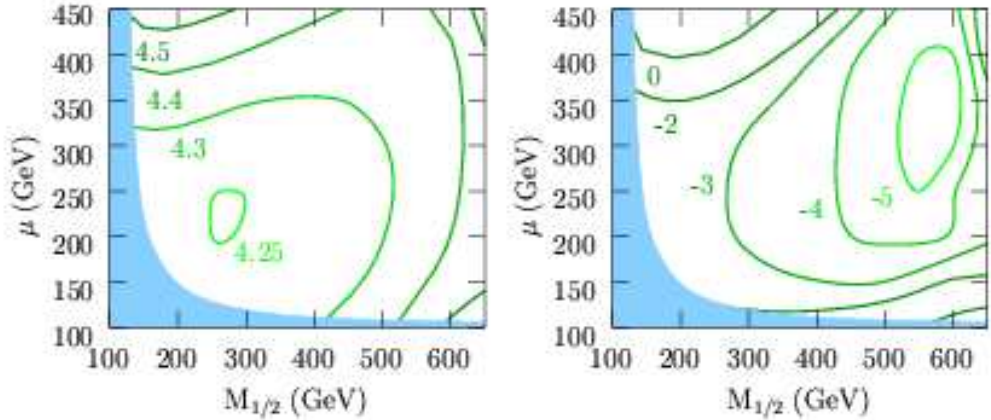}} \caption{Contours
of constant $m_b(m_b)$[GeV] (Left) and $\delta m_b$ in \% (Right)
for $m_{16} = 2000$ GeV.} \label{fig:fig2}
\end{figure}

{\em The bottom line result of our analysis is that Yukawa
unification is possible only in a narrow region of SUSY parameter
space.} The result is also easy to understand.  Since for $\mu >
0$ and $\delta m_b/ m_b \leq - 2\%$ we need $|\Delta m_b^{\tilde
\chi}|
> \Delta m_b^{\tilde \chi}$.   However $ \Delta m_b^{\tilde g}
\approx \frac{2 \alpha_3}{3 \pi} \; \frac{\mu m_{\tilde
g}}{m_{\tilde b}^2} \; tan\beta $, $\;\;  \Delta m_b^{\tilde
\chi^+} \approx \frac{\lambda_t^2}{16 \pi^2} \; \frac{\mu
A_t}{m_{\tilde t}^2} \; tan\beta $ and $\Delta m_b^{\log} \approx
\frac{\alpha_3}{4 \pi} \log(\frac{\tilde m^2}{M_Z^2}) \sim 6$ \%.
In order to enhance the chargino contribution, we can make the
numerator larger by making $A_t$ large and negative.   This is
accomplished by making $A_0$ at $M_G$ large and negative, i.e.
$A_t << 0 \Longleftrightarrow  A_0 << 0$.   This also has the
effect of making the denominator for the chargino contribution
smaller since the stop mass matrix is of the form
$\left(\begin{array}{cc}
m_{\tilde t}^2 &  m_t \; A_t \\
m_t \; A_t  &   m_{\tilde {\bar t}}^2 \end{array} \right)$.  As a
consequence we naturally obtain $ m_{\tilde t} << m_{\tilde b} $;
enhancing the chargino, in comparison to the gluino, contribution.
Of course in order not to have a negative stop mass squared we
need to make $m_{16}$ large. As a result of the $\chi^2$ analysis
we find that good fits require $A_0 \sim - 2 \; m_{16}, \;\;$
$m_{10} \sim \sqrt{2} \; m_{16},\;\;$ $m_{16} \geq 2  \; {\rm TeV}
\gg \mu, \ M_{1/2},\;\;$ and $\Delta m_H^2 \sim 10$ \%.   In
Fig.~\ref{fig:fig1} we show the $\chi^2$ contours for two
different values of $m_{16}$ as a function of $\mu$ and $M_{1/2}$.
It is clear that $\chi^2$ improves as we increase $m_{16}$.  Note
also that the dominant pull for $\chi^2$ is due to the bottom
quark mass corrections as can be seen in Fig.~\ref{fig:fig2}.

\section{Summary -- Minimal SO(10) SUSY Model}
Before discussing some phenomenological consequences of the
MSO$_{10}$SM, let us summarize the main ingredients of the model.
We assume a supersymmetric SO(10) GUT with quarks and leptons in
{\bf 16}s.   In addition we assume that the minimal Higgs content
of the MSSM ($H_u, \ H_d$) are contained in a single {\bf 10}.
Finally for the third family we assume the minimal Yukawa
interaction with  $$W \supset \lambda \ {\bf 16_3 \ 10_H \ 16_3}
.$$  The direct consequences of MSO$_{10}$SM follow.
\begin{itemize}
\item {\em Gauge coupling unification} -- $ \alpha_G, \ M_G, \
\epsilon_3 \sim - 4 \%$;

\item  {\em Yukawa unification} --  $ \lambda_t \ = \ \lambda_b \
= \ \lambda_\tau \ = \ \lambda_{\bar \nu_\tau}  \ \equiv \ \lambda
$;

\item   {\em Soft SUSY breaking parameters}\footnote{In general
$m_{16}$ would have a family index.  We have made the additional
assumption of a universal squark and slepton mass.} -- $m_{16}, \
m_{10},\ A_0, \ M_{1/2}, \ \tan\beta \approx 50, \ \Delta m_H^2$;

\item {\em Satisfying} -- $A_0 \sim - 2 \; m_{16}, \;\;$ $m_{10}
\sim \sqrt{2} \; m_{16}, \;\;$ $m_{16}  \geq 2  \; {\rm TeV} \gg
\mu, \ M_{1/2}, \;\;$ and $\Delta m_H^2 \sim 10 \%$.
\end{itemize}
The last condition is required in order to fit the precision low
energy electroweak data, including the top, bottom and tau masses.
In addition to the above defining properties of the MSO$_{10}$SM,
we find two additional direct consequences of the model.   The
first is a ``natural" inverted scalar mass hierarchy which
ameliorates the SUSY flavor and CP problems.  Secondly, the rates
for proton decay due to dimension 5 operators are decreased.  We
discuss these two unexpected benefits below.

\subsection{Inverted Scalar Mass Hierarchy}

One way to ameliorate the SUSY flavor and CP problems is to demand
that the 1st \& 2nd generation squarks and sleptons are heavy with
mass $>>$ TeV, while the 3rd generation scalars are light with
mass $\leq$ TeV.  This is easily seen by focusing on the most
severe flavor and CP violating processes~\cite{gabbiani}.   The
best bounds are for processes involving the two lightest families.
For example, we have~\cite{gabbiani} --

\begin{itemize}

\item $B(\mu \rightarrow e \gamma) < 1.2 \times 10^{-11}$
$\Longrightarrow$  $|(\delta_{12}^l)_{LL}| < 2.1 \times 10^{-3}
(m_{\tilde l}({\rm GeV})/100)^2$
\medskip

or $|(\delta_{12}^l)_{LL}| <  0.8 \ (m_{\tilde l}({\rm
TeV})/2)^2$;

\item  $\Delta m_{K} < $ Exp. $\Longrightarrow$
$\sqrt{|Re(\delta_{12}^d)^2_{LL}|} < 1.9 \times 10^{-2} (m_{\tilde
q}({\rm GeV})/500)$
\medskip

or $\sqrt{|Re(\delta_{12}^d)^2_{LL}|} < 7.6 \times 10^{-2} \
(m_{\tilde q}({\rm TeV})/2)$;

\item  $\epsilon_{K} < $ Exp. $\Longrightarrow$
$\sqrt{|Im(\delta_{12}^d)^2_{LL}|} < 1.5 \times 10^{-3} (m_{\tilde
q}({\rm GeV})/500)$
\medskip

or $\sqrt{|Im(\delta_{12}^d)^2_{LL}|} < 6.0 \times 10^{-3} \
(m_{\tilde q}({\rm TeV})/2)$;

\item $d_N^e \sim 2 (100 /m_{\tilde l}({\rm GeV}))^2 sin\Phi_{A,B}
\times 10^{-23} {\rm e \ cm} < 4.3 \times 10^{-27} {\rm e \ cm} $
$\Longrightarrow$ $sin\Phi_{A,B}  < 4 \times 10^{-4} \times
(m_{\tilde l}({\rm GeV})/100)^2$
\medskip

or $sin\Phi_{A,B}  < 0.16 \times (m_{\tilde l} \ ({\rm
TeV})/2)^2$.
\end{itemize}  Although a significant degeneracy of the first and
second generation squarks and sleptons is still required, it does
not require serious fine tuning.  In fact, the flavor and CP
problems are now completely amenable to solutions using
non-abelian family symmetries.   The question one now faces is how
to obtain an inverted scalar mass hierarchy with the ratio of
scalar masses $S$ satisfying $S \equiv \tilde m_{1,2}^2/\tilde
m_3^2 \gg 1$.

One way of obtaining this inverted scalar mass hierarchy is to
assume that it results from Planck/GUT scale physics. However, it
was shown that an inverted scalar mass hierarchy can be generated
``naturally" as a consequence of renormalization group
running~\cite{crunching}.   This latter possibility requires
specific soft SUSY breaking boundary conditions at $M_G$.  In
particular it was found that the following boundary conditions can
lead to values of $S \geq 400$~\cite{crunching}.  Surprisingly,
{\em these boundary conditions are the same required by Yukawa
unification.}

\begin{itemize} \item $ m_Q^2 \ = \ m_U^2 \ = \ m_D^2 \ = \ m_L^2 \ = \ m_E^2  \
\equiv m_{16}^2 $;

\item $ A_t \ = \ A_b \ = \ A_\tau \ \equiv \ A_0$;

\item $ M_1 \ = \ M_2 \ = \ M_3  \ \equiv \ M_{1/2} $;

\item $ m_{H_u} \ = \  m_{H_d} \ \equiv \ m_{10}$; and

\item  $A_0^2 \ = \ 2 \ m_{10}^2 \ = \ 4 \ m_{16}^2$ with $m_{16}
>> 1$ TeV.
\end{itemize}

\subsection{Suppressing proton decay}

Nucleon decay rates are significantly constrained by data from
Super-Kamiokande~\cite{protondecay}.   In particular the decay
mode $p \rightarrow K^+ + \bar \nu_\tau$, due to dimension 5
operators, is typically the dominant decay mode.   In the large
$\tan\beta$ regime the dominant Feynman diagram is given below.

This one loop integral results in a loop factor characteristically
of order  $${\rm Loop \;\; Factor} = \frac{\lambda_t \;
\lambda_\tau}{16 \pi^2} \frac{\sqrt{\mu^2 + M_{1/2}^2}}{m_{16}^2}
.$$   Note the loop factor is minimized in the limit $\mu, M_{1/2}
\;\; \ll \;\; m_{16}$.   This limit is once again consistent with
Yukawa unification.   Moreover it is only consistent with
radiative EWSB with split $H_u, \; H_d$ masses.

\begin{center}
\SetPFont{Helvetica}{15}
%\SetScale{0.8}

\begin{picture}(300,200)(0,0)

\SetColor{Magenta} \ArrowLine(40,160)(100,160)
\PText(30,160)(0)[rc]{s}

\SetColor{Black} \ArrowLine(40,40)(100,40) \Text(30,40)[1c]{$\nu$}
\Text(37,33)[1c]{$\tau$}

\SetColor{Blue} \ArrowLine(220,160)(160,100)
\ArrowLine(220,40)(160,100) \PText(230,40)(0)[lc]{d}
\PText(234,164)(0)[rc]{u}

\SetColor{Brown} \DashArrowLine(100,160)(160,100){5}
\DashArrowLine(100,40)(160,100){5} \PText(130,150)(0)[lc]{t}
\Photon(130,156)(138,158){2}{1} \Text(130,50)[lc]{$\tau$}
\Photon(130,57)(138,59){2}{1}

\SetColor{Green} \Line(100,40)(100,160)
\Photon(100,40)(100,160){7}{4.5} \PText(80,100)(0)[rc]{H}
\Photon(72,105)(80,107){2}{1}

\SetColor{Magenta} \ArrowLine(130,180)(40,180)
\PText(35,190)(0)[rc]{u}

\SetColor{Blue} \ArrowLine(220,180)(130,180)
\PText(225,190)(0)[lc]{u}

\SetColor{Red} \Vertex(160,100){3}

\SetColor{Black}
\end{picture}
\end{center}

\section{Phenomenology}

Let us now consider some predictions of the MSO$_{10}$SM.

\subsection{Light Higgs mass}
First consider the light Higgs mass.   In the MSSM the light Higgs
mass has an upper bound of order 130 GeV.  This upper limit is
achieved for large $\tan\beta$.  Moreover the large radiative
corrections to the Higgs mass are dominated by heavy stop masses.
In our case we have $\tan\beta \sim 50$, however we have
relatively light stop, sbottom, and stau masses.   As a result we
find~\cite{bdr}

$$ m_h = 114 \pm 5 \pm 3 \;\; {\rm GeV} .$$

\begin{figure}[htbp]
\epsfxsize=15cm \centerline{\epsfbox{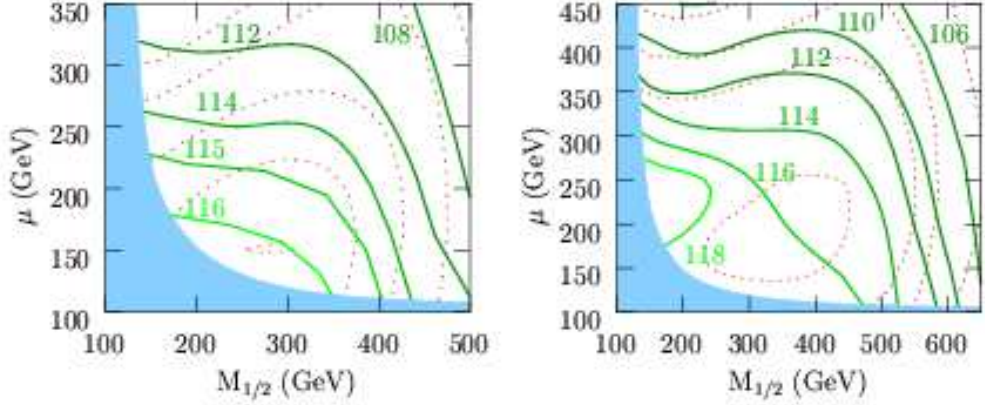}} \caption{Contours
of constant $m_h$ [GeV] (solid lines) with $\chi^2$ contours from
Fig 1 (dotted lines) for $m_{16} = 1500$ GeV (Left) and $m_{16} =
2000$ GeV (Right).} \label{fig:fig3}
\end{figure}
 In Fig. \ref{fig:fig3} we show the light Higgs
mass contours as a function of $\mu, \ M_{1/2}$ for two values of
$m_{16}$.   For a more detailed analysis of the light Higgs mass
prediction, see~\cite{bdr}.

\subsection{Muon anomalous magnetic moment}
The anomalous magnetic moment of the muon scales as $(\mu \
M_{1/2} \ \tan\beta)/m_{16}^4$.   Since we have $m_{16} \geq 2
\;\; {\rm TeV} $, we find~\cite{bdr}

$$ a_\mu^{SUSY} \leq 6 \times 10^{-10} .$$

\subsection{$\tilde \chi^0$ LSP --  Dark Matter}

When $m_{16}$ is large, the standard neutralino annihilation
channels via squark/slepton exchange diagrams are severely
suppressed.  This typically leads to an excess cosmological
abundance of neutralinos with $\Omega_\chi h^2 \gg 0.3$.  However,
in the large $\tan\beta$ regime the neutralino annihilation
channel $( \tilde \chi^0 \tilde \chi^0 \rightarrow A^0 \rightarrow
$ hadrons ) is significantly enhanced.   In fact, this
annihilation channel is so effective that, on resonance,
$\Omega_\chi h^2 \ll 0.01$.    Thus we find, on the sides of the
broad resonance peak, cosmological abundances consistent with dark
matter observations~\cite{drrr}.  In Fig. \ref{fig:darkmatter} we
present an analysis of dark matter abundances in the MSO$_{10}$SM.
The green band is the region with acceptable values of
$\Omega_\chi h^2$.   Note that we have also included contours of
constant branching ratio $B(B_s \rightarrow \mu^+ \mu^-)$.   This
is important since this process is extremely sensitive to the
value of the CP odd Higgs mass $m_A$.

\subsection{Large $\tan\beta$ and Quark Flavor Violation}

\begin{figure}[htbp]
\epsfxsize=8cm \centerline{\epsfbox{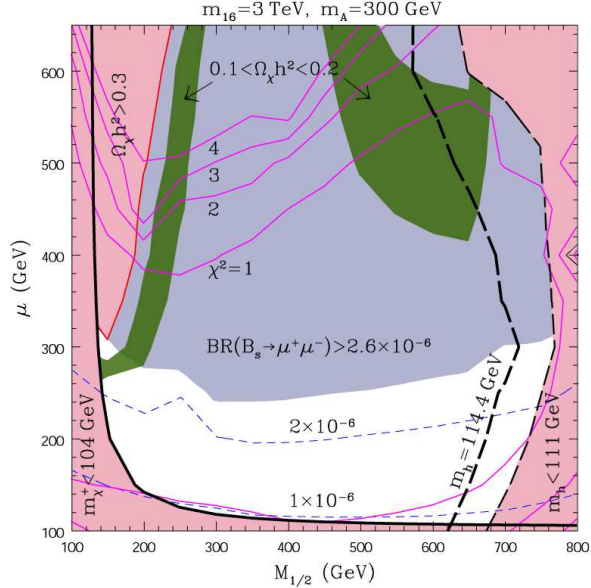}}
\caption{Contours of constant $\chi^2$ for $m_{16} = 3$ TeV and
$m_{A} = 300$ GeV. The red regions are excluded by
$m_{\chi^+}<104$ GeV (below and to the left of a black solid
curve), $m_{h}<111$ GeV (on the right) and by $\Omega_\chi
h^2>0.3$. To the right of the black broken line one has
$m_{h}<114.4$ GeV. The green band corresponds to the preferred
range $0.1<\Omega_\chi h^2<0.2$, while the white regions below
(above) it correspond to $\Omega_\chi h^2<0.1$ ($0.2<\Omega_\chi
h^2<0.3$).  Also marked are contours of constant ${\rm \bf
BR}(B_s\rightarrow \mu^+ \ \mu^-)$. The blue regions in the lower
two panels are excluded by ${\rm \bf BR}(B_s\rightarrow \mu^+\
\mu^-)>2.6\times10^{-6}$.} \label{fig:darkmatter}
\end{figure}

It has been shown that in the large $\tan\beta$ regime there are
significant one loop SUSY threshold corrections to CKM matrix
elements~\cite{bpr}.  Once these corrections are included in an
effective two Higgs doublet model below the SUSY breaking scale,
the Higgs couplings are no longer flavor
diagonal~\cite{pokorski,bsmumu}.   Hence the process $B_s
\rightarrow \mu^+ \mu^-$ can proceed through s-channel CP odd
Higgs exchange with a $\tan\beta$ enhanced branching ratio $B(B_s
\rightarrow \mu^+ \mu^-) \propto \tan\beta^4$.    The effective
two Higgs doublet Yukawa coupling to down quarks is given below.
The matrices $\lambda_{di}^{diag}, \ \Delta \lambda_d^{ij} \
(\delta \lambda_d)$ are the zeroth order down quark Yukawa
coupling in a diagonal basis and the one loop correction to the
Higgs couplings due to gluino (chargino) loops.

\begin{eqnarray}
{\cal {L}}_{eff}^{ddH}= &  -\bar{d}_{Li} \ \lambda_{di}^{diag} \
d_{Ri} \  H_d^{0*} & \nonumber \\ & -\bar{d}_{Li} \ \Delta
\lambda_d^{ij} \ d_{Rj} \ H_d^{0*} & \nonumber \\ & -\bar{d}_{Li}
\ \delta \lambda_d^{ij} \ d_{Rj} \ H_u^0 + {\rm h.c.} . &
\nonumber
\end{eqnarray}
As a result of the chargino loop correction, which is proportional
to the square of the up quark Yukawa matrix, we must
re-diagonalize the down quark mass matrix
\begin{eqnarray}
m_{d}^{Diagonal} &=& V_d^L \ \left[\lambda_d^{diag} +\Delta
\lambda_d + \delta \lambda_d \tan\beta  \right] \ V_d^{R\dagger} \
\frac{v\cos\beta}{\sqrt{2}} .\nonumber \end{eqnarray} For large
values of $\tan\beta$, this results in a significant correction to
the CKM matrix~\cite{bpr}.  We then obtain the following couplings
to the neutral Higgs mass eigenstates $h, \ H,\ A$ given
by~\cite{pokorski,bsmumu}.
\begin{eqnarray}
{\cal {L}}_{FV}^{i\neq j} =& -\frac{1}{\sqrt{2}}\bar{d'}_i \left[
F^h_{ij} \ P_R+ F^{h*}_{ji} \ P_L \right] d'_j \ h & \nonumber  \\
& -\frac{1}{\sqrt{2}}\bar{d'}_i \left[ F^H_{ij} \ P_R +
F^{H*}_{ji} \ P_L \right] d'_j \ H &
\nonumber \\
& -\frac{i}{\sqrt{2}}\bar{d'}_i \left[ F^A_{ij} \ P_R +
F^{A*}_{ji} \ P_L \right] d'_j \ A, & {\rm where} \nonumber
\end{eqnarray}
\begin{eqnarray}
F^h_{ij} &\simeq& \delta \lambda_d^{ij}  (1+\tan^2\beta) \cos\beta \ \cos(\alpha-\beta), \nonumber \\
F^H_{ij} &\simeq& \delta \lambda_d^{ij}  (1+\tan^2\beta) \cos\beta \ \sin(\alpha-\beta), \nonumber \\
F^A_{ij} &\simeq& \delta \lambda_d^{ij} (1+\tan^2\beta) \
\cos\beta . \nonumber
\end{eqnarray}
It is the flavor violating coupling $F^A_{2 3}$ which gives the
direct $B_s \ A^0$ coupling~\cite{bsmumu}.    Note, the branching
ratio $B(B_s \rightarrow \mu^+ \mu^-)$ in the cosmologically
allowed region is close to the CDF bound (see Fig.
\ref{fig:darkmatter}).   In \cite{drrr} we show that the process
$B_s \rightarrow \mu^+ \mu^-$ may soon be observed.

\section{Two loose ends}
\subsection{Fine tuning ?}

We have been considering large squark and slepton masses with
$m_{16} \geq 2$ TeV.  Since we have a ``natural" inverted scalar
mass hierarchy, the third generation squarks and sleptons are
typically lighter than a TeV.   Hence the radiative corrections to
the Higgs mass, in the effective low energy theory, are not large.
For example the radiative corrections at the electroweak scale are
of order $\delta m_h^2 \propto   \frac{\lambda_\tau^2}{16 \pi^2} \
m_{\tilde \tau}^2$ and they are safe for  $m_{\tilde \tau} \leq 1$
TeV.

However, there is still the question of whether radiative EWSB
requires significant fine tuning.   It has been shown that the $Z$
mass is most sensitive to the value of the gluino mass,
$M_3$~\cite{Kane:2002ap}.   For example, with $\tan\beta = 35$,
the following relation was obtained~\cite{Kane:2002ap}

$$M_Z^2 = - 1.5 \ \mu^2 \ + \ 5.0 \ M_{1/2}^2 \ + \ 0.2 \ A_0^2 \ + \ 1.5 \
m_{16}^2 \ - \ 1.2 \ m_{H_u}^2  \ - \ 0.08 \ m_{H_d}^2 \ + \
\cdots .$$  BUT, recall we have $\mu, \ M_{1/2} \ll m_{16}$.  Thus
this problem is ameliorated somewhat, although it is not
completely eliminated. (See also the talk by S. Pokorski, this
conference.)

Finally, as discussed earlier the fine tuning for EWSB in the
regime of large $\tan\beta$ is of order 1/$\tan\beta$ when one has
Higgs mass splitting~\cite{rattazzi}.

\subsection{SUSY Breaking Mechanism ?}

We have found that Yukawa unification in the MSO$_{10}$SM is only
consistent with the low energy data in a narrow region of soft
SUSY breaking parameter space.  It is clear that this idea would
be considerably strengthened if there was a mechanism which
``naturally" broke SUSY in this way.   Unfortunately, this is not
the case for the known SUSY breaking mechanisms.

For example, gauge mediated SUSY breaking [GMSB] gives $A_0 = 0$
at the messenger scale.  This is {\em bad}.  Yukawa deflected GMSB
can have non-zero $A_0$ proportional to Yukawa couplings. Perhaps
this might work, however the standard gauge contribution would
have to be suppressed.  Moduli-dominated string SUSY breaking may
be possible~\cite{crunching}.  The generic formula for stringy
SUSY breaking is given by~\cite{Brignole:1995fb}

\begin{eqnarray}
m^2_\alpha = & (1 + 3 \overrightarrow{n_\alpha} \cdot
\overrightarrow{\Theta^2}) \ m^2_{3/2} & \nonumber \\
M_{1/2} \sim & 0 & \nonumber \\
A_{\alpha \beta \gamma} = & \pm \sqrt{3} [1 + n^i_\alpha + n^i_\beta + n^i_\gamma - Y^i_{\alpha \beta \gamma}] \
\Theta^i \ m_{3/2} & \nonumber
\end{eqnarray}
where  $n_\alpha$ is the modular weight of the field $\alpha$;
$T_i, \ (i = 1, \cdots, 6)$ are moduli; $\Theta_i$ parametrize the
direction of the goldstino in the $T_i$ field space, and
$Y^i_{\alpha \beta \gamma} = 2 (Re T_i) \
\partial_{T_i} \ln h_{\alpha \beta \gamma}$ where  $h_{\alpha \beta \gamma}$
are the dimensionless Yukawa couplings.  With the following
modular weights~\cite{crunching}: $n_{Q,U,D,L,E,N} = (0, -1/2,
-1/2, 0,0,0)$; $ n_{H_u, H_d} = (-1/2, -1/2, 0, 0,0,0) $, and
assuming $(\Theta^2)^i =  (0, 0, 1/3, 2/3 , 0,0) $, and
$Y^i_{\alpha \beta \gamma} \sim 0$, one finds $m_{3/2}^2 = 2
m_{16}^2$ and $$ A_0 = \pm 2 m_{16}, \;\;\;  m_{10} = \sqrt{2}
m_{16}.$$  The only problem with this idea is the absence of any
existing string model with these properties.  Thus the problem of
a ``natural" SUSY breaking mechanism consistent with MSO$_{10}$SM
Yukawa unification is the most urgent open theoretical question
requiring further work.

\section{Summary}

In this talk I have defined the minimal SO(10) SUSY model and
discussed some of its phenomenological consequences.   The model
predicts:

$\bullet$ Gauge coupling unification with  $\alpha_G, \ M_G, \
\epsilon_3 \sim - 4 \%$;

$\bullet$ Yukawa unification with $ \lambda_t \ = \ \lambda_b \ =
\ \lambda_\tau \ = \ \lambda_{\bar \nu_\tau}  \ \equiv \ \lambda$,
and

$\bullet$ Soft SUSY breaking parameters given by~\footnote{A
universal soft breaking scalar mass $m_{16}$ for each family of
squarks and sleptons is guaranteed by SO(10).  However, we have
assumed in addition that $m_{16}$ is family independent.} $m_{16},
\ m_{10},\ A_0, \ M_{1/2}, \ \tan\beta, \ \Delta m_H^2$.

As a result of a $\chi^2$ analysis~\cite{bdr} we find that the low
energy precision electroweak data, including the top, bottom and
tau masses, only gives good fits for soft SUSY breaking parameters
satisfying:

$\bullet$ $A_0 \sim - 2 \; m_{16}$, $\;\; m_{10} \sim  \sqrt{2} \;
m_{16}$, $\;\; m_{16}  \geq 2  \; {\rm TeV} \gg \mu, \ M_{1/2} \
$, and $\; \Delta m_H^2 \sim 10 \%$.

This region of parameter space has the virtue of giving:

$\bullet$ a ``natural" inverted scalar mass hierarchy which
ameliorates the SUSY flavor and CP problems, and in addition

$\bullet$ suppresses proton decay via dimension 5 operators.
\medskip

The MSO$_{10}$SM makes the following predictions: \begin{itemize}
\item It gives $\tan\beta \sim 50$ and a light stop.  As a
consequence we find~\cite{bdr} $m_h = 114 \pm 5 \pm 3$ GeV;

\item The decay $B_s \rightarrow \mu^+ \ \mu^-$ is enhanced and
may be observable in the near future~\cite{bsmumu};

\item  The SUSY contribution to the muon anomalous magnetic moment
is suppressed with $a_\mu^{SUSY} < 6 \times 10^{-10}$~\cite{bdr};
and

\item  Finally, it gives cosmologically acceptable abundances of
neutralino dark matter~\cite{drrr} (see Roszkowski, this
 conference).
 \end{itemize}

 \noindent
 {\bf Acknowledgements}
 I would like to acknowledge the support of the Institute for
 Advanced Study (and a grant in aid from the Monell Foundation)
 where this talk was written and also partial support
 from the U.S. Department of Energy grant \# DOE/ER/01545-844.  I
 would also like to thank the organizers of PASCOS'03 for their
 kind hospitality during my stay in Mumbai.


\begin{thebibliography}{99}
\bibitem{bdr}  T. Bla\v{z}ek, R. Derm\' \i \v sek and S. Raby,
\prl{88}, 111804 (2002); \prd{65}, 115004 (2002).

\bibitem{brt}   T. Bla\v{z}ek, S. Raby and K. Tobe, \prd{60}, 113001 (1999);
ibid., \prd{62}, 055001 (2000); R. Derm\' \i \v sek and S. Raby,
\prd{62}, 015007 (2000).

\bibitem{rattazzi} R. Rattazzi and U. Sarid, \prd{53}, 1553 (1996).

\bibitem{ewsb}  M. Olechowski and S. Pokorski, \plb{344},  201  (1995);
 D. Matalliotakis and H.P. Nilles,  \npb{435}, 115 (1995);
N. Polonsky and A. Pomarol, \prd{51}, 6532 (1995); H. Murayama, M.
Olechowski and S. Pokorski,  \plb{371}, 57 (1996).

\bibitem{susygut1} S. Dimopoulos, S. Raby and F. Wilczek, \prd{24}, 1681 (1981).

\bibitem{susygut2}  S. Dimopoulos and H. Georgi,
\npb{193}, 150 (1981), L.E. Ibanez and G.G. Ross, \plb{105}, 439
(1981); N. Sakai, Z. Phys. {\bf C11}, 153 (1981); M.B. Einhorn and
D.R. Jones, \npb{196}, 475 (1982); W.J. Marciano and G.
Senjanovic, \prd{25}, 3092 (1982).

\bibitem{susygutexp} U. Amaldi, W. de Boer and H. F\"urstenau, \plb{260}, 447
(1991);  J. Ellis, S. Kelly and D.V. Nanopoulos,  \plb{260}, 131
(1991); P. Langacker and M. Luo, \prd{44}, 817 (1991).

\bibitem{so10} T. Banks, \npb{303}, 172 (1988);
M. Olechowski and S. Pokorski, \plb{214}, 393 (1988); S. Pokorski,
\npb{13} (Proc. Supp.), 606 (1990); B. Ananthanarayan, G.
Lazarides and Q. Shafi, \prd{44}, 1613 (1991); Q. Shafi and B.
Ananthanarayan, ICTP Summer School lectures (1991); S. Dimopoulos,
L.J. Hall and S. Raby, \prl{68}, 1984 (1992), \prd{45}, 4192
(1992); G. Anderson et al., \prd{47}, 3702 (1993); B.
Ananthanarayan, G. Lazarides and Q. Shafi, \plb{300}, 245 (1993);
G. Anderson et al.,  \prd{49}, 3660 (1994); B. Ananthanarayan, Q.
Shafi and X.M. Wang, \prd{50}, 5980 (1994).



\bibitem{tobe}
K.~Tobe and J.~D.~Wells, arXiv:hep-ph/0301015;  D.~Auto, H.~Baer,
C.~Balazs, A.~Belyaev, J.~Ferrandis and X.~Tata,
arXiv:hep-ph/0302155.

\bibitem{threshcorr}  L.J. Hall, R. Rattazzi and U. Sarid, \prd{50}, 7048 (1994); R. Hempfling,
\prd{49}, 6168 (1994),  M. Carena et al., \npb{426}, 269 (1994);
R. Rattazzi and U. Sarid, \prd{53}, 1553 (1996);  D. Pierce et
al., \npb{491}, 3 (1997).

\bibitem{bpr} T. Blazek, S.
Pokorski and S. Raby, \prd{52}, 4151 (1995).

\bibitem{gminus2} G.W. Bennett et al. (Muon (g-2) Collaboration),
\prl{89} (2002) 101804; Erratum-ibid., {\bf 89} (2002) 129903;
M.~Davier, S.~Eidelman, A.~Hocker and Z.~Zhang,
arXiv:hep-ph/0208177.

\bibitem{nath}  U. Chattopadhyay and P. Nath, \prd{65}, 075009 (2002); S. Komine and M. Yamaguchi, hep-ph/0110032.

\bibitem{gabbiani} Gabbiani, Gabrieli, Masiero \& Silvestrini, {\it
Nucl. Phys.}{\bf B477}, 321 (1996).

\bibitem{crunching} Bagger, Feng, Polonsky, \& Zhang,  {\it Phys.
Lett.} {\bf B473}, 264 (2000).

\bibitem{protondecay} V. Lucas and S. Raby, \prd{55}, 6986 (1997); T. Goto, T.  Nihei, \prd{59}, 115009 (1999);
 K.S. Babu and M.J. Strassler, hep-ph/9808447; H. Murayama and A. Pierce, \prd{65}, 055009 (2002).

\bibitem{drrr} R. Derm\' \i \v sek, S. Raby, L. Roszkowski and R. Ruiz de
 Austri, in preparation.  See also talk by L. Roszkowski, this
 conference.

\bibitem{pokorski}
P.~H.~Chankowski and S.~Pokorski, arXiv:hep-ph/9707497.

\bibitem{bsmumu} G. Isidori and A. Retico, hep-ph/0110121;
A. Dedes, H.K. Dreiner and U. Nierste, hep-ph/0108037; K.S. Babu
and C. Kolda, \prl{84}, 228 (2000) and references therein.

\bibitem{Kane:2002ap}
G.~L.~Kane, J.~Lykken, B.~D.~Nelson and L.~T.~Wang, Phys.\ Lett.\
B {\bf 551}, 146 (2003).

\bibitem{Brignole:1995fb}
A.~Brignole, L.~E.~Ibanez, C.~Munoz and C.~Scheich, Z.\ Phys.\ C
{\bf 74}, 157 (1997).




\end{thebibliography}
\end{document}